\documentclass[reqno,11pt]{amsart}
\usepackage{multind}
\usepackage{hyperref}
\usepackage{graphicx}
\usepackage{amscd}
\usepackage{slashed}
\usepackage[mathscr]{eucal}

\input amssym.def
\input amssym.tex

\def\goth{\frak}
\def\double{\mathbb}

\def\cc{{\double C}}

\def\rr{{\double R}}
\def\zz{{\double Z}}

\def\da{{\partial}_{\!\mbox{\tiny A}}}
\def\db{{\partial}_{\mbox{\tiny B}}}
\def\dD{{\partial}_{\mbox{\tiny D}}}
\def\ddd{{/\!\!\!\partial}}
\def\DDD{{/\!\!\!\!D}}
\def\dda{{/\!\!\!\partial}_{\!\!\mbox{\tiny A}}}
\def\ddb{{/\!\!\!\partial}_{\!\mbox{\tiny B}}}
\def\ddD{{/\!\!\!\partial}_{\!\mbox{\tiny D}}}

\def\ot{\otimes}
\def\op{\oplus}

\def\bb{\begin{eqnarray}}
\def\ee{\end{eqnarray}}

\newcommand{\beq}{\begin{equation}}
\newcommand{\eeq}{\end{equation}}

\newtheorem{definition}{Definition}[section]

\newtheorem{proposition}{Proposition}[section]

\newcommand{\Proof}{\begin{proof}}
\newcommand{\QED}{\end{proof} \noindent}

\title[DOp's of simple type and the EH-action with cosmological constant]{The Einstein-Hilbert action with cosmological constant as a functional of generic form}

\author[J.\ Tolksdorf]{J\"urgen Tolksdorf \\ \\ \today}
\address{Max Planck Institute for Mathematics in the Sciences, Leipzig, Germany}
\email{Juergen.Tolksdorf@mis.mpg.de}
\thanks{The research leading to these results has received funding from the
European Research Council under the European Union's Seventh Framework
Programme (FP7/2007-2013) / ERC grant agreement n$^\circ$~267087.}

\begin{document}
\maketitle

\begin{abstract}
The geometrical underpinnings of a specific class of Dirac operators is discussed. It is demonstrated
how this class of Dirac operators allow to relate various geometrical functionals like, for example, the 
Yang-Mills action and the functional of non-linear $\sigma-$models (i.e. of (Dirac) harmonic maps).  
These functionals are shown to be similar to the Einstein-Hilbert action with cosmological constant (EHC). The EHC may thus be regarded as a ``generic functional''.  As a byproduct, the geometrical setup presented also allows to avoid the issue of ``fermion doubling'' as usually encountered, for instance,
in the geometrical discussion of the Standard Model in terms of Dirac operators. Furthermore, it is
demonstrated how the geometrical setup presented allows to derive the cosmological constant term
of the EHC from the Einstein-Hilbert functional and the action of a purely gauge coupling Higgs field.
\end{abstract}

\tableofcontents



\section{Introduction}
The Einstein-Hilbert action with a cosmological constant (EHC), see \eqref{ehc} below, is known to 
be the most general functional whose Euler-Lagrange equation yields a (geometrically) divergency free tensor field that can be build from the metric and its first and second derivatives, only (see Section 3). 
In contrast to the pure Einstein-Hilbert action:
\beq
\label{eh}
\mathcal{I}_{\mbox{\tiny EH}}(g_{\mbox{\tiny M}}) := \int_M\ast scal(g_{\mbox{\tiny M}})\,,
\eeq
which can be expressed in terms of ``quantized Clifford connections'' (for the notation and terminology used, please see the next section), the EHC can be expressed in terms of ``Dirac operators of simple 
type''. This class of Dirac operators provides a natural generalization of quantized Clifford connections 
in the sense that the Bochner-Laplacian associated with a Dirac operator is still defined in terms of a 
Clifford connection.  We demonstrate that the EHC has a ``generic form'' when expressed in terms of 
Dirac operators of simple type. We discuss the functional of non-linear $\sigma-$models
and the Yang-Mills action from this point of view. Though these functionals yield rather different Euler-Lagrange equations, both functionals may nonetheless be recast into a form similar to the EHC. For 
this a certain class of Clifford module bundles is introduced. This class of bundles also allows to avoid 
the ``doubling of fermions'' needed to geometrically describe the Standard Model action in terms of Dirac operators (see \cite{Tolksdorf} and the References sited therein).

Dirac operators of simple type describe the dynamics of the Standard Model fermions. In loc site 
it has been shown that also the bosonic functional of the Standard Model can be described by Dirac operators of simple type. As a premise for this, however, it is necessary to assume a specific bi-module structure of the underlying Clifford module bundle. Moreover, one has to double the Clifford module bundle to introduce curvature terms into Dirac operators. In the Standard Model fermions are described in terms of spinors. But twisted spinor bundles as discussed, for instance, in \cite{Tolksdorf} do not permit a bi-module structure necessary to describe the Standard Model action in terms of Dirac operators of simple type. Without a bi-module structure, however, the pure ``kinetic term'' of the Higgs sector of the Standard Model cannot be described in terms of the setup introduced in loc. site. For the same reason the geometrical setup discussed in \cite{Tolksdorf} cannot be used to describe functionals like the functional of Dirac harmonic maps (non-linear $\sigma-$models). Also, in order to describe Yang-Mills gauge theory within the geometrical scheme presented in loc site one has to use a class of Dirac operators (of ``Pauli type''), which do not form a distinguished subset of the set of all Dirac operators. This is different to the case of simple type Dirac operators.

The main motivation of the present paper is to remedy these flaws and to demonstrate that Dirac operators of simple type actually provide a ``generic root'' of a variety of various seemingly different functionals, including non-linear $\sigma-$models and the Yang-Mills action, as well as the purely gauge coupling
Higgs field.  All of these functionals can be re-cast into the form of Einstein's ``biggest blunder''\footnote{Actually, there seems no written text in which Einstein himself called his introduction of the cosmological constant ``the biggest blunder of my life''. See, however, Ref. \cite{Gamov}.} \cite{Misner et al}\label{REF}. 
We also demonstrate how the geometrical scheme of Dirac operators of simple type provides a purely
geometrical relation between the cosmological constant and the kinetic term of the Higgs action.

\section{The geometrical setup and basic definitions}
In this section we introduce the basic geometrical setup and fix the notation used. For sake
of self-consistency we recapitulate some basic facts about Dirac operators acting on sections of general Clifford module bundles.

In the sequel, $(M,g_{\mbox{\tiny M}})$ always denotes a smooth orientable (semi-)Riemannian
manifold of finite dimension $n\equiv p + q$. The index of the (semi-)Riemannian metric 
$g_{\mbox{\tiny M}}$ is $s\equiv p - q\not\equiv 1\text{~mod~}4$. The {\it bundle of exterior forms} of degree $k\geq 0$ is 
denoted by $\Lambda^{\!k}T^\ast\!M\rightarrow M$ with its canonical projection. Accordingly, the
{\it Grassmann bundle} is given by 
$\Lambda T^\ast\!M \equiv \bigoplus_{k\geq 0}\Lambda^{\!k}T^\ast\!M\rightarrow M$. It naturally
inherits a metric denoted by $g_{\mbox{\tiny$\Lambda{\rm M}$}}$, such that the direct sum is orthogonal 
and the restriction of $g_{\mbox{\tiny$\Lambda{\rm M}$}}$ to degree one equals to the fiber metric 
$g^\ast_{\mbox{\tiny M}}$ of the cotangent bundle $T^\ast\!M\rightarrow M$.

%In what follows we restrict to the case where $p - q \not\equiv 1~{\rm mod~} 4$, such that the 
%Clifford algebra $Cl_{p,q}$ of the quadratic space $\rr^{p,q}$ is {\it simple} (see, for instance, 
%\cite{Lounesto})\label{REF}.

The bundle of (complexified) {\it Clifford algebras} is denoted by $Cl_{\mbox{\tiny M}}\rightarrow M$, where, again, we do not explicitly mention its canonical projection. As a vector bundle the Clifford bundle is canonically isomorphic to the Grassmann bundle (see below). Accordingly, the Clifford bundle also inherits a natural metric structure, such that its restriction to the generating sub-space $T^\ast\!M\subset Cl_{\mbox{\tiny M}}$ again reduces to $g^\ast_{\mbox{\tiny M}}$. In what follows, the
Grassmann and the Clifford bundle are mainly regarded as complex bundles, though we do not explicitly indicate their complexification. Accordingly, all (linear) maps are understood as complex linear extensions 
of the underlaying real linear maps.

The mutually inverse ``musical isomorphisms'' in terms of $g_{\mbox{\tiny M}}$ 
(resp. $g^\ast_{\mbox{\tiny M}}$) are denoted by $\phantom{x}^{\flat/\sharp}:\,TM\simeq T^\ast\!M$, such that, for instance, $g_{\mbox{\tiny M}}(u,v) = g^\ast_{\mbox{\tiny M}}(u^\flat,v^\flat)$ for all 
$u,v\in TM$.

A smooth complex vector bundle $\pi_{\mbox{\tiny$\mathcal{E}$}}:\,\mathcal{E}\rightarrow M$ is called a
{\it Clifford module bundle}, provided there is a {\it Clifford map}. That is, there is a smooth linear (bundle) map (over the identity on $M$)
\beq
\label{cliffmap}
\begin{split}
\gamma_{\mbox{\tiny$\mathcal{E}$}}:\,T^\ast\!M &\longrightarrow {\rm End}(\mathcal{E})\\
\alpha &\mapsto \gamma_{\mbox{\tiny$\mathcal{E}$}}(\alpha)\,,
\end{split}
\eeq
satisfying $\gamma_{\mbox{\tiny$\mathcal{E}$}}(\alpha)^2 = 
\epsilon g^\ast_{\mbox{\tiny M}}(\alpha,\alpha){\rm Id}_{\mbox{\tiny$\mathcal{E}$}}$. Here,
$\epsilon\in\{\pm 1\}$ depends on how the {\it Clifford product} is defined. That is,
$\alpha^2 := \pm g^\ast_{\mbox{\tiny M}}(\alpha,\alpha)1_{\mbox{\tiny Cl}}\in Cl_{\mbox{\tiny M}}$,
for all $\alpha\in T^\ast\!M\subset Cl_{\mbox{\tiny M}}$ and $1_{\mbox{\tiny Cl}}\in Cl_{\mbox{\tiny M}}$ 
denotes the unit element. 

A Clifford map \eqref{cliffmap} is known to induce a unique homomorphism 
$\Gamma_{\!\mbox{\tiny$\mathcal{E}$}}:\,Cl_{\mbox{\tiny M}}\rightarrow{\rm End}(\mathcal{E})$ of associative algebras with unit, 
such that $\Gamma_{\!\mbox{\tiny$\mathcal{E}$}}(\alpha) = \gamma_{\mbox{\tiny$\mathcal{E}$}}(\alpha)$, for all $\alpha\in T^\ast\!M\subset Cl_{\mbox{\tiny M}}$. 
To explicitly mention the underlying structure we denote a Clifford module bundle also by
\beq
\label{cliffmodbdl}
\begin{split}
\pi_{\mbox{\tiny$\mathcal{E}$}}:\,(\mathcal{E},\gamma_{\mbox{\tiny$\mathcal{E}$}}) &\longrightarrow 
(M,g_{\mbox{\tiny M}})\\
z &\mapsto x = \pi_{\mbox{\tiny$\mathcal{E}$}}(z)\,.
\end{split}
\eeq

If the Clifford module bundle is $\zz_2-$graded, with grading involution being given by
$\tau_{\mbox{\tiny$\mathcal{E}$}}\in{\rm End}(\mathcal{E})$, then 
the Clifford map $\gamma_{\mbox{\tiny$\mathcal{E}$}}$ is assumed to be odd: 
$\gamma_{\mbox{\tiny$\mathcal{E}$}}(\alpha)\tau_{\mbox{\tiny$\mathcal{E}$}} 
= -\tau_{\mbox{\tiny$\mathcal{E}$}}\gamma_{\mbox{\tiny$\mathcal{E}$}}(\alpha)$, for all 
$\alpha\in T^\ast\!M$. Furthermore, if the vector bundle is supposed to be hermitian, with the hermitian product being denoted by $\big<\cdot,\cdot\big>_{\!\mbox{\tiny$\mathcal{E}$}}$, then the Clifford map and the grading involution are supposed to bei either hermitian or anti-hermitian. In this case, we call 
\eqref{cliffmodbdl} an {\it odd hermitian Clifford module bundle}. Notice that in the sequel the term ``hermitian'' does not necessarily imply that $\big<\cdot,\cdot\big>_{\!\mbox{\tiny$\mathcal{E}$}}$ is supposed to be positive definite. In fact, the signature of the fiber metric may depend on the signature
of $g_{\mbox{\tiny M}}$ as, for example, in the case of the Clifford bundle associated to 
$(M,g_{\mbox{\tiny M}})$.

The sheaf of sections of any bundle $\pi_{\mbox{\tiny$\mathcal{W}$}}:\,\mathcal{W}\rightarrow M$ is denoted by $\goth{S}ec(M,W)$. In the particular case of the cotangent bundle, however, we follow the common notation and denote the corresponding sheaf of sections by 
$\Omega(M)\equiv\goth{S}ec(M,\Lambda T^\ast\!M)$. Accordingly,
$\Omega^k(M,{\rm End}(\mathcal{E}))\equiv
\goth{S}ec(M,\Lambda^{\!k}T^\ast\!M\ot_{\mbox{\tiny M}}{\rm End}(\mathcal{E}))$ are the 
``${\rm End}(\mathcal{E})-$valued forms'' of degree $k\geq 0$.

From the Wedderburn structure theorems about invariant algebras one infers that 
(see \cite{Atiyah et al}, \cite{Berline et al} and \cite{Greub})\label{REF}
\beq
\label{enddecomp}
{\rm End}(\mathcal{E}) \simeq Cl_{\mbox{\tiny M}}\ot_{\mbox{\tiny M}}{\rm End}_\gamma(\mathcal{E})\,,
\eeq
where ${\rm End}_\gamma(\mathcal{E})\subset{\rm End}(\mathcal{E})$ denotes the sub-algebra of
endomorphisms on \eqref{cliffmodbdl} which commute with the Clifford action provided by 
$\gamma_{\mbox{\tiny$\mathcal{E}$}}$.

As a consequence,
\beq
\Omega^0(M,{\rm End}(\mathcal{E})) \simeq \Omega(M,{\rm End}_\gamma(\mathcal{E}) \equiv
\bigoplus_{k\geq 0}\Omega^k(M,{\rm End}_\gamma(\mathcal{E})\,.
\eeq

%On a ($\zz_2-$graded, hermitian) Clifford module bundle \eqref{cliffmodbdl} there is a 
%{\it canonical one-form} $\Theta\in\Omega^1(M,{\rm End}^-(\mathcal{E}))$, given by
%\beq
%\Theta(v) := \frac{\epsilon}{n}\gamma(v^\flat)\quad(v\in TM)\,.
%\eeq
%Here, the two mutually inverse mappings:
%$\hspace{-.3cm}\phantom{X}^{\flat/\sharp}: TM\simeq T^\ast\!M$, denote the ``musical isomorphisms'' between the tangent and the cotangent bundle of $M$ with respect to $g_{\mbox{\tiny M}}$. That is,
%$g^\ast_{\mbox{\tiny M}}(\alpha,\beta) = g_{\mbox{\tiny M}}(\alpha^\sharp,\beta^\sharp)$ and
%$g^\ast_{\mbox{\tiny M}}(u^\flat,v^\flat) = g_{\mbox{\tiny M}}(u,v)$, for all $\alpha,\beta\in T^\ast\!M$ and
%$u,v\in TM$ (over the same base point $x\in M$).

The linear map
\beq
\label{quantmap}
\begin{split}
\delta_\gamma:\,\Omega(M,{\rm End}(\mathcal{E})) &\longrightarrow 
\Omega^0(M,{\rm End}(\mathcal{E}))\\
\alpha\otimes\goth{B} &\mapsto ~{/\!\!\!\!\alpha}\goth{B} \equiv
\gamma_{\mbox{\tiny$\mathcal{E}$}}\big(\sigma_{\!\mbox{\tiny Ch}}^{-1}(\alpha)\big)\goth{B}
\end{split}
\eeq
is called the {\it ``quantization map''}. It is determined by the linear isomorphism called {\it symbol map}:
\beq
\label{symbolmap}
\begin{split}
\sigma_{\!\mbox{\tiny Ch}}:\,Cl_{\mbox{\tiny M}} &\stackrel{\simeq}{\longrightarrow} \Lambda T^\ast\!M\\
\goth{a} &\mapsto \Gamma_{\!\mbox{\tiny Ch}}(\goth{a})1_{\mbox{\tiny$\Lambda$}}\,.
\end{split}
\eeq
Here, $1_{\mbox{\tiny$\Lambda$}}\in\Lambda T^\ast\!M$ is the unit element. The homomorphism 
$\Gamma_{\!\mbox{\tiny Ch}}:\,Cl_{\mbox{\tiny M}}\rightarrow{\rm End}(\Lambda T^\ast M)$ 
is given by the canonical Clifford map:
\beq
\begin{split}
\gamma_{\mbox{\tiny Cl}}:\,T^\ast\!M &\longrightarrow {\rm End}(\Lambda T^\ast\!M)\\
v &\mapsto \left\{
  \begin{array}{ccc}
    \Lambda T^\ast\!M & \longrightarrow & \hspace{-2cm}\Lambda T^\ast\!M \\ 
    \omega & \mapsto & \epsilon int(v)\omega + ext(v^\flat)\omega\,,
  \end{array}
\right.
\end{split}
\eeq
where, respectively, $``int"$ and $``ext"$ indicate ``interior'' and ``exterior'' multiplication.

\begin{definition}\label{cliffextension}
A Clifford module bundle $\pi_{\mbox{\tiny$\mathcal{E'}$}}:\,
(\mathcal{E}',\gamma_{\mbox{\tiny$\mathcal{E'}$}})\longrightarrow(M,g_{\mbox{\tiny M}})$ is called an 
``extension'' of the Clifford module bundle \eqref{cliffmodbdl}, provided there is a bundle embedding
$\iota:\,\mathcal{E}\hookrightarrow\mathcal{E}'$ (over the identity on $M$), such that for all 
$\alpha\in T^\ast\!M$ and $z\in\mathcal{E}$:
\beq
\gamma_{\mbox{\tiny$\mathcal{E'}$}}(\alpha)\iota(z) = \iota(\gamma_{\mbox{\tiny$\mathcal{E}$}}(\alpha)z)\,.
\eeq

Furthermore, in the case of odd hermitian Clifford module bundles one assumes that
\beq
\begin{split}
\tau_{\mbox{\tiny$\mathcal{E'}$}}\iota(z) &= \iota(\tau_{\mbox{\tiny$\mathcal{E}$}}z)\,,\\
\big<\iota(z_1),\iota(z_2)\big>_{\!\mbox{\tiny$\mathcal{E'}$}} &= 
\big<z_1,z_2\big>_{\!\mbox{\tiny$\mathcal{E}$}}\,,
\end{split}
\eeq
for all $z,z_1,z_2\in\mathcal{E}$.
\end{definition}

\begin{definition}
A smooth vector bundle 
$\pi_{\mbox{\tiny$\mathcal{E}$}}:\,\mathcal{E}\longrightarrow(M,g_{\mbox{\tiny M}})$ is called a 
``Clifford bi-module bundle'', if there are Clifford maps:
$\gamma_{\mbox{\tiny$\mathcal{E}$}},\,\gamma'_{\mbox{\tiny$\mathcal{E}$}}:\,
T^\ast\!M\longrightarrow{\rm End}(\mathcal{E})$, such that for all $\alpha,\beta\in T^\ast\!M$:
\beq
\gamma_{\mbox{\tiny$\mathcal{E}$}}(\alpha)\gamma'_{\mbox{\tiny$\mathcal{E}$}}(\beta) =
\gamma'_{\mbox{\tiny$\mathcal{E}$}}(\beta)\gamma_{\mbox{\tiny$\mathcal{E}$}}(\alpha)\,.
\eeq
\end{definition}

Every Clifford module bundle \eqref{cliffmodbdl} possesses a natural extension to a Clifford bi-module
bundle, which is given by
\beq
\label{twistedcliffmodbdl}
\begin{split}
\iota:\,\mathcal{E} &\hookrightarrow \mathcal{E'} := \mathcal{E}\ot_{\mbox{\tiny M}}Cl_{\mbox{\tiny M}}\\
z &\mapsto z \equiv z\ot 1_{\mbox{\tiny Cl}}\,.
\end{split}
\eeq

The extension \eqref{twistedcliffmodbdl} also provides an extension of odd hermitian Clifford module bundles with respect to the grading involution and hermitian structure, respectively
\beq
\begin{split}
\tau_{\mbox{\tiny$\mathcal{E'}$}} &:= \tau_{\mbox{\tiny$\mathcal{E}$}}\ot{\rm Id}_{\mbox{\tiny Cl}}\,,\\
\big<\cdot,\cdot\big>_{\!\mbox{\tiny$\mathcal{E'}$}} &:=
\big<\cdot,\cdot\big>_{\!\mbox{\tiny$\mathcal{E}$}}\big<\cdot,\cdot\big>_{\!\mbox{\tiny Cl}}\,.
\end{split}
\eeq
Here, $\big<\cdot,\cdot\big>_{\!\mbox{\tiny Cl}}$ denotes the hermitian structure that is defined in terms 
of the symbol map \eqref{symbolmap} and the canonical extension of $g^\ast_{\mbox{\tiny M}}$ to the Grassmann bundle.

Notice that the extension \eqref{twistedcliffmodbdl} completely fixes the Clifford action to be given by 
$\gamma_{\mbox{\tiny$\mathcal{E}'$}} = 
\gamma_{\mbox{\tiny$\mathcal{E}$}}\!\ot{\rm Id}_{\mbox{\tiny Cl}}$. We call the canonical bi-module extension \eqref{twistedcliffmodbdl} the {\it Clifford twist} of the Clifford module bundle \eqref{cliffmodbdl}.

Similar to \eqref{enddecomp}, one has
\beq
{\rm End}(\mathcal{E'}) \simeq 
Cl_{\mbox{\tiny M}}\ot_{\mbox{\tiny M}}{\rm End}_\gamma(\mathcal{E})\ot_{\mbox{\tiny M}} 
\!\big(Cl_{\mbox{\tiny M}}\ot_{\mbox{\tiny M}} Cl^{\mbox{\tiny op}}_{\mbox{\tiny M}}\big)\,,
\eeq
where $Cl^{\mbox{\tiny op}}_{\mbox{\tiny M}}\rightarrow M$ is the bundle of {\it opposite Clifford algebras}.

We call in mind that a {\it Dirac operator} $\DDD$ is a first order differential operator acting on sections 
$\psi\in\goth{S}ec(M,\mathcal{E})$, such that
$[\,\DDD,f]\psi = \gamma_{\mbox{\tiny$\mathcal{E}$}}(df)\psi$ for all smooth functions 
$f\in\mathcal{C}^\infty(M)$.
The set of all Dirac operators on \eqref{cliffmodbdl} is denoted by 
$\goth{Dir}(\mathcal{E},\gamma_{\mbox{\tiny$\mathcal{E}$}})$. It provides an affine space over the vector space $\Omega^0(M,{\rm End}(\mathcal{E}))$. Moreover, on odd Clifford module bundles Dirac operators are {\it odd} operators, i.e. 
$\,\DDD\tau_{\mbox{\tiny$\mathcal{E}$}} = -\tau_{\mbox{\tiny$\mathcal{E}$}}\,\DDD$.

We call the Dirac operator ${/\!\!\!\!\nabla}^{\mbox{\tiny$\mathcal{E}$}}\equiv
\delta_\gamma(\nabla^{\mbox{\tiny$\mathcal{E}$}})$ the ``quantization'' of a connection 
$\nabla^{\mbox{\tiny$\mathcal{E}$}}$ on \eqref{cliffmodbdl}. Let
$e_1,\ldots,e_n\in\goth{S}ec(U,TM)$ be a local frame and $e^1,\ldots,e^n\in\goth{S}ec(U,T^\ast\!M)$
its dual frame. For $\psi\in\goth{S}ec(M,\mathcal{E})$ one obtains
\beq
{/\!\!\!\!\nabla}^{\mbox{\tiny$\mathcal{E}$}}\!\psi := 
\sum_{k = 1}^n\delta_\gamma(e^k)\nabla_{\!\!e_k}^{\mbox{\tiny$\mathcal{E}$}}\psi =
\sum_{k = 1}^n\gamma_{\mbox{\tiny$\mathcal{E}$}}(e^k)\nabla_{\!\!e_k}^{\mbox{\tiny$\mathcal{E}$}}\psi\,,
\eeq
where the natural embedding $\Omega(M)\hookrightarrow\Omega(M,{\rm End}(\mathcal{E})),\;\omega\mapsto \omega\equiv\omega\ot{\rm Id}_{\mbox{\tiny$\mathcal{E}$}}$ is taken into account.

Every Dirac operator has a canonical first-order decomposition:
\beq
\label{1dirdecomp}
\DDD = \ddb + \Phi_{\mbox{\tiny D}}\,.
\eeq
Here, $\db$ denotes the {\it Bochner connection} on \eqref{cliffmodbdl}, that is defined by $\,\DDD$ as
\beq
\label{bochcon}
2ev_g\big(df,\db\psi\big) := \epsilon\big([\,\DDD^2,f] - \delta_{\!g} df\big)\psi\qquad
\big(\psi\in\goth{S}ec(M,\mathcal{E})\big)\,,
\eeq
with $ev_g$'' being the evaluation map with respect to $g_{\mbox{\tiny M}}$ and
$\delta_{\!g}$ the dual of the exterior derivative (see \cite{Berline et al}).
 
The zero-order section $\Phi_{\mbox{\tiny D}} := \,\DDD - \ddb\in\goth{S}ec(M,{\rm End}(\mathcal{E}))$ is thus uniquely determined by $\,\DDD$. We call the Dirac operator $\ddb$ the {\it``quantized Bochner connection''}. 

A (linear) connection on 
\eqref{cliffmodbdl} is called a {\it Clifford connection} if the corresponding covariant derivative 
$\nabla^{\mbox{\tiny$\mathcal{E}$}}$ ``commutes'' with the Clifford map 
$\gamma_{\mbox{\tiny$\mathcal{E}$}}$ in the following sense:
\beq
\label{cliffcon1}
[\nabla^{\mbox{\tiny$\mathcal{E}$}}_{\!\!X},\gamma_{\mbox{\tiny$\mathcal{E}$}}(\alpha)] = 
\gamma_{\mbox{\tiny$\mathcal{E}$}}\big(\nabla^{\mbox{\tiny$T^\ast\!M$}}_{\!\!X}\alpha\big)\qquad
\big(X\in\goth{S}ec(M,TM),\;\alpha\in\goth{S}ec(M,T^\ast\!M)\big)\,,
\eeq
where $\nabla^{\mbox{\tiny$T^\ast\!M$}}$ denotes the Levi-Civita connection on the co-tangent bundle
of $M$ with respect to $g_{\mbox{\tiny M}}^\ast$.

We denote Clifford connections as $\da$ since a Clifford connection is seen to be parametrized 
by a family of locally defined forms $A\in\Omega^1(U,{\rm End}_\gamma(\mathcal{E}))$. This basically
follows from \eqref{enddecomp}. Therefore, Clifford connections certainly provide a distinguished class of connections on a Clifford module bundle.

Since $\goth{Dir}(\mathcal{E},\gamma_{\mbox{\tiny$\mathcal{E}$}})$ is an affine space, every Dirac operator can be written as
\beq
\DDD = \dda + \Phi\,.
\eeq
However, this decomposition is far from being unique, as opposed to the first-order decomposition \eqref{1dirdecomp}. In particular, the
section $\Phi\in\goth{S}ec(M,{\rm End}(\mathcal{E}))$ depends on the chosen Clifford connection $\da$. 
In general, a Dirac operator does not uniquely determine a Clifford connection.

\begin{definition}
A Dirac operator is said to be of ``simple type'' provided that $\Phi_{\!\mbox{\tiny D}}$ anti-commutes
with the Clifford action:
\beq
\Phi_{\!\mbox{\tiny D}}\gamma_{\mbox{\tiny$\mathcal{E}$}}(\alpha) = 
-\gamma_{\mbox{\tiny$\mathcal{E}$}}(\alpha)\Phi_{\!\mbox{\tiny D}}\qquad\big(\alpha\in T^\ast\!M\big)\,.
\eeq
\end{definition}

It follows that a Dirac operator of simple type uniquely determines a Clifford connection $\da$ together with a zero-order operator $\phi_{\mbox{\tiny D}}\in\goth{S}ec(M,{\rm End}_\gamma(\mathcal{E}))$, such that
(c.f. \cite{Tolksdorf})\label{REF}
\beq
\label{simpletypedop}
\DDD = \dda + \tau_{\mbox{\tiny$\mathcal{E}$}}\phi_{\mbox{\tiny D}}\,.
\eeq
These Dirac operators play a basic role in the geometrical description of the Standard Model
(c.f. \cite{Tolksdorf})\label{REF}. They are also used in the context of the family index theorem (see, for instance, \cite{Berline et al})\label{REF}. 

Apparently, Dirac operators of simple type provide a natural generalization of quantized Clifford connections. They are distinguished (and fully characterized!) by the fact that they build the most general class of Dirac operators with the property that their Bochner connections \eqref{bochcon} are also Clifford connections (see the next section).

Notice that $\phi_{\mbox{\tiny D}}\in\goth{S}ec(M,{\rm End}^-_\gamma(\mathcal{E}))$ in the case of odd Clifford module bundles, with ${\rm End}^\pm(\mathcal{E})\subset{\rm End}(\mathcal{E})$ denoting the sub-algebras of even and odd endomorphisms.

\section{The EHC action as a generic functional}
In appropriate physical units, the {\it Einstein-Hilbert functional} (action of gravity) with a cosmological constant added is given by
\beq
\label{ehc}
\mathcal{I}_{\mbox{\tiny EHC}} := \int_M\!\!\ast\big(scal(g_{\mbox{\tiny M}}) + \Lambda\big)\,.
\eeq
Here,  ``$\ast$'' denotes the Hodge map with respect to $g_{\mbox{\tiny M}}$ and a chosen orientation
of $M$. The smooth function $scal(g_{\mbox{\tiny M}})\in\mathcal{C}^\infty(M)$ is the scalar curvature
and $\Lambda\in\rr$ denotes the {\it cosmological constant}. 

In fact, due to {\it Lovelock's Theorem}, the structure of the functional $\mathcal{I}_{\mbox{\tiny EHC}}$ 
is basically unique (for $dim(M) = 4$) if one requires that the Euler-Lagrange equations for the metric are
of second order, tensorial and have vanishing divergency 
%uniquely determined by the requirement that the Euler-Lagrange equation is tensorial and only 
%depends on the metric and its first and second derivative 
(c.f. \cite{Lovelock} and the editors remark H2 on page 285 in 
\cite{Pauli}; for a refinement of this statement in terms of ``natural geometry'' we refer to 
\cite{Navarro et al}).\label{REF}

We demonstrate that the functional $\mathcal{I}_{\mbox{\tiny EHC}}$ has a generic form 
in the sense that several other geometrical functionals may be recast into the form similar to \eqref{ehc}. As an example, we present in this vein the functional of non-linear $\sigma-$models (Dirac harmonic maps) 
and the Yang-Mills action.

On a given Clifford module bundle \eqref{cliffmodbdl} every Dirac operator naturally defines two connections: The Bochner connection \eqref{bochcon} and the {\it ``Dirac connection''}
\beq
\label{dircon}
\dD := \db + \omega_{\mbox{\tiny D}}\,.
\eeq
Here, $\omega_{\mbox{\tiny D}}\equiv\Theta\Phi_{\mbox{\tiny D}}\in\Omega^1(M,{\rm End}(\mathcal{E}))$ 
is the {\it ``Dirac form''}, with $\Theta(v) := \frac{\epsilon}{n}\gamma_{\mbox{\tiny$\mathcal{E}$}}(v^\flat)$ being the {\it canonical one-form} for all $v\in TM$. It is the right-inverse of the quantization map \eqref{quantmap} restricted to $\Omega^1(M,{\rm End}(\mathcal{E}))$. The canonical one-form also plays a
basic role in the construction of twister operators. In terms of the canonical one-form Clifford connections may be characterized as follows: A connection on a Clifford module bundle \eqref{cliffmodbdl} is a Clifford connection if and only if it leaves the canonical one-form covariantly constant:
\beq
\label{cliffcon2}
\nabla^{\mbox{\tiny$T^\ast\!M\!\ot\!{\rm End}(\mathcal{E})$}}_{\!\!X}\Theta \equiv 0\qquad
\big(X\in{\goth S}ec(M,TM)\big)\,.
\eeq

On a Clifford module bundle with a chosen Dirac operator:
\beq
\pi_{\mbox{\tiny$\mathcal{E}$}}:\,
(\mathcal{E},\gamma_{\mbox{\tiny$\mathcal{E}$}},\,\DDD)\longrightarrow(M,g_{\mbox{\tiny M}})\,,
\eeq 
the Dirac connection is distinguished for it is uniquely determined by $\,\DDD$. Also, the Dirac connection has the property that $\ddD\equiv\delta_\gamma(\dD) = \DDD$. Notice that neither $\ddb = \,\DDD$, nor
are $\db$ and $\dD$ are Clifford connections, in general.

Every Dirac operator is known to have a unique {\it second order decomposition}
\beq
\label{2nddirdecomp}
\DDD^2 = \triangle_{\mbox{\tiny B}} + V_{\mbox{\tiny D}}\,,
\eeq
where the {\it Bochner-Laplacian} (or ``trace Laplacian'') is given in terms of the Bochner connection as
$\triangle_{\mbox{\tiny B}} := \epsilon ev_g\big(\db^{\mbox{\tiny$T^\ast\!M\!\ot\!\mathcal{E}$}}\circ\db\big)$.
The trace of the zero-order operator $V_{\mbox{\tiny D}}\in\goth{S}ec(M,{\rm End}(\mathcal{E}))$ explicitly reads (c.f. \cite{Tolksdorf}): \label{REF}
\beq
\label{trdirpot}
tr_{\!\mbox{\tiny$\mathcal{E}$}}V_{\!\mbox{\tiny D}} = 
tr_{\!\gamma}\!\left(curv(\,\DDD) - \varepsilon\,{\rm ev}_{\!g}(\omega_{\mbox{\tiny D}}^2)\right) - 
\varepsilon\delta_{\!g}\big(tr_{\!\mbox{\tiny$\mathcal{E}$}}\omega_{\mbox{\tiny D}}\big)\,,
\eeq
where $curv(\,\DDD)\in\Omega^2(M,{\rm End}(\mathcal{E}))$ denotes the curvature of the Dirac 
connection of $\,\DDD\in\goth{Dir}(\mathcal{E},\gamma_{\mbox{\tiny$\mathcal{E}$}})$ and
$tr_{\!\gamma} := tr_{\!\mbox{\tiny$\mathcal{E}$}}\circ\delta_{\gamma}$ the ``quantized trace''.

Let $M$ be closed compact. We call the functional
\beq
\label{diract}
\begin{split}
\mathcal{I}_{\mbox{\tiny D}}:\, \goth{Dir}(\mathcal{E},\gamma_{\mbox{\tiny$\mathcal{E}$}}) &\rightarrow\cc\\
\DDD &\mapsto \int_M \ast tr_{\!\mbox{\tiny$\mathcal{E}$}}V_{\!\mbox{\tiny D}}
\end{split}
\eeq
the {\it ``universal Dirac action''} and
\beq
\label{totdiract}
\begin{split}
\mathcal{I}_{\mbox{\tiny D,tot}}:\, \goth{Dir}(\mathcal{E},\gamma_{\mbox{\tiny$\mathcal{E}$}})\times
\goth{S}ec(M,\mathcal{E}) &\rightarrow\cc\\
(\,\DDD,\psi) &\mapsto \int_M \ast\big(\big<\psi,\,\DDD\psi\big>_{\!\mbox{\tiny$\mathcal{E}$}} +
tr_{\!\mbox{\tiny$\mathcal{E}$}}V_{\!\mbox{\tiny D}}\big)
\end{split}
\eeq
the {\it ``total Dirac action''}. Here, ``$\ast$'' is the Hodge map with respect to $g_{\mbox{\tiny M}}$
and a chosen orientation of $M$.

If the Dirac connection of $\,\DDD$ is a Clifford connection, then $\dD = \db$. In this case, the Dirac action
\eqref{diract} reduces to the Einstein-Hilbert functional \eqref{eh}.

In contrast, for Dirac operators of simple type the Dirac action becomes 
\beq
\label{stypediract}
\mathcal{I}_{\mbox{\tiny D}}\big(\dda + \tau_{\mbox{\tiny$\mathcal{E}$}}\phi_{\mbox{\tiny D}}\big) = 
\int_M\!\ast\big(-\mbox{\small$\epsilon\frac{rk(\mathcal{E})}{4}$}scal(g_{\mbox{\tiny M}}) + 
tr_{\!\mbox{\tiny$\mathcal{E}$}}\phi_{\mbox{\tiny D}}^2\big)\,,
\eeq
with $rk(\mathcal{E})\geq 1$ being the rank of \eqref{cliffmodbdl}. This is a direct consequence of 
Lemma 4.1 and the Corollary 4.1 of Ref. \cite{Tolksdorf} (see also Sec. 6 in loc site).\label{REF}

Apparently, the restriction of the Dirac action \eqref{diract} to Dirac operators of simple type 
\eqref{simpletypedop} formally coincides with the Einstein-Hilbert action \eqref{ehc} with a 
cosmological constant, where (up to numerical factors)
\beq
\Lambda = tr_{\!\mbox{\tiny$\mathcal{E}$}}\phi_{\mbox{\tiny D}}^2 \equiv \pm\|\phi_{\mbox{\tiny D}}\|^2\,.
\eeq

Similar to \eqref{ehc}, the Einstein equation of \eqref{stypediract} yields $\|\phi_{\mbox{\tiny D}}\| = const.$ 
as long as the section $\phi_{\mbox{\tiny D}}\in\goth{S}ec(M,{\rm End}^-_\gamma(\mathcal{E}))$ does not depend on $g_{\mbox{\tiny M}}$.  In the case of a transitive action, this reduces the gauge group 
$\mathcal{G} \equiv\goth{S}ec(M,{\rm Aut}_\gamma(\mathcal{E}))$ to the stabilizer group of a chosen
point on the (hyper) sphere $\|\phi_{\mbox{\tiny D}}\| = const.$ and therefore spontaneously breaks the (gauge) symmetry that is provided by the structure of the underlying Clifford module bundle 
\eqref{cliffmodbdl}. This reduction of the gauge group is in complete analogy to the symmetry breaking induced by the Higgs potential of the Standard Model (we refer to \cite{Tolksdorf}, for a more thorough discussion of this point).\label{REF}

To proceed let $\chi\in\Omega(M,{\rm End}^-_{\gamma'}(\mathcal{E}'))$. With respect to a local 
(oriented) orthonormal basis $e^1,\ldots,e^n\in\goth{S}ec(U,T^\ast\!M)$ we may write 
$(U\subset M,\text{open})$:
\beq
\begin{split}
\chi &\stackrel{loc.}{=} \sum_{k=0}^n\;\sum_{1\leq i_1<i_2<\cdots<i_k\leq n}
\hspace{-.5cm}e^{i_1}\wedge e^{i_2}\wedge\cdots\wedge e^{i_k}\ot\chi_{i_1i_2\cdots i_k}\\
&\equiv
\sum_{I}e^I\ot\chi_I\,.
\end{split}
\eeq
We put
\beq
%{/\!\!\!\!\chi} 
\phi_{\mbox{\tiny D}} := \sum_I\chi_I e^I\in
\goth{S}ec(M,{\rm End}^-_{\gamma'}(\mathcal{E}'))\,,
\eeq
where for $\chi_I = \sum_{J,K} \varphi_{\mbox{\tiny$IJK$}}\ot\goth{a}_J\ot\goth{b}_K\in
{\rm End}_{\gamma'}(\mathcal{E}')
= 
{\rm End}^-_\gamma(\mathcal{E})\ot Cl_{\mbox{\tiny M}}\ot Cl^{\mbox{\tiny op}}_{\mbox{\tiny M}}$:
\beq
\chi_I e^I \equiv 
\sum_{J,K} \varphi_{\mbox{\tiny$IJK$}}\ot\goth{a}_J\sigma_{\mbox{\tiny Ch}}^{-1}(e^I)\ot\goth{b}_K\,.
\eeq
The explicit form of the coefficients $\varphi_{\mbox{\tiny$IJK$}}\in{\rm End}_\gamma(\mathcal{E})$ is related to the structure of the Clifford module bundle \eqref{cliffmodbdl}. This structure may yield a 
metric dependent ``cosmological constant''
\beq
\Lambda = \pm\|\,\phi_{\mbox{\tiny D}}\|^2\,.
\eeq
The EHC then gives rise to additional first order constraints on the fields 
$\varphi_{\mbox{\tiny$IJK$}}$ via the condition of a vanishing divergency of the associated 
energy-momentum current.

As an application we discuss the functional of Dirac harmonic maps, the Yang-Mills action and a combination of both as special cases of \eqref{stypediract}.

\section{The functional of non-linear $\sigma-$models}
In this section we specify the Clifford module bundle \eqref{cliffmodbdl} and Dirac operators of 
simple type, such that the total Dirac action reduces to the functional of Dirac-Harmonic maps.
We also discuss the ''energy functional'' of geodesics within this geometrical setup.

For $k = 1,2$ let $\pi_k:\,(\mathcal{E}_k,\gamma_k)\rightarrow(M_k,g_k)$ be odd hermitian Clifford 
module bundles of rank $N_k\geq 1$ over smooth orientable (semi-)Riemannian manifolds of dimensions 
$n_k = p_k + q_k\equiv dim(M_k)$ and signatures $s_k = p_k - q_k \in \zz$. The corresponding Clifford 
bundles are denoted by $Cl_k\rightarrow(M_k,g_k)$. The grading involution and hermitian products 
read $\tau_k$ and $\big<\cdot,\cdot\big>_{\!k}$, respectively. In the sequel we assume $M_1$ to be closed compact. 

Let $\varphi:\,M_1\rightarrow M_2$ be a smooth map. We set
\beq
\label{cliffharmap}
\pi_{\mbox{\tiny$\mathcal{E}$}}:\,\mathcal{E} := 
\mathcal{E}_1\ot_{\mbox{\tiny{$M_1$}}}\varphi^\ast\mathcal{E}_2
\longrightarrow (M_1,g_1)\,,
\eeq
as well as
\beq
\begin{split}
\gamma_{\mbox{\tiny$\mathcal{E}$}} &:= 
\gamma_1\ot{\rm Id}_{\mbox{\tiny$\varphi^{\!\ast}\!\mathcal{E}_2$}}\,,\\
\tau_{\mbox{\tiny$\mathcal{E}$}} &:= \tau_1\ot\tau_2|_\varphi\,,\\
\big<\cdot,\cdot\big>_{\!\mbox{\tiny$\mathcal{E}$}} &:= 
\big<\cdot,\cdot\big>_{\!1}\big<\cdot,\cdot\big>_{\!2}|_\varphi\,.
\end{split}
\eeq
That is, in the case considered the Clifford module bundle \eqref{cliffmodbdl} is a {\it twisted Clifford 
module bundle} with the twisting provided by the pull-back of the odd hermitian vector bundle 
$\pi_2:\,\mathcal{E}_2\rightarrow M_2$ with respect to the smooth map $\varphi$.

Finally, we set
\beq
\label{clifftwistharmap}
\pi_{\mbox{\tiny$\mathcal{E}'$}}:\,\mathcal{E}'\ := \mathcal{E}\ot_{\mbox{\tiny$M_1$}}Cl_1
\longrightarrow (M_1,g_1)
\eeq
and
\beq
\begin{split}
\gamma_{\mbox{\tiny$\mathcal{E}'$}} &:= 
\gamma_{\mbox{\tiny$\mathcal{E}$}}\ot{\rm Id}_{\mbox{\tiny${\rm Cl}_1$}}\,,\\
\tau_{\mbox{\tiny$\mathcal{E}'$}} &:= 
\tau_{\mbox{\tiny$\mathcal{E}$}}\ot{\rm Id}_{\mbox{\tiny${\rm Cl}_1$}}\,,\\
\big<\cdot,\cdot\big>_{\!\mbox{\tiny$\mathcal{E}'$}} &:= 
\big<\cdot,\cdot\big>_{\!\mbox{\tiny$\mathcal{E}$}}\big<\cdot,\cdot\big>_{\!\mbox{\tiny${\rm Cl}_1$}}\,.
\end{split}
\eeq
That is, the odd hermitian Clifford module bundle
$\pi_{\mbox{\tiny$\mathcal{E}'$}}:\,(\mathcal{E}',\gamma_{\mbox{\tiny$\mathcal{E}'$}})\rightarrow(M_1,g_1)$ is the Clifford twist of the odd hermitian Clifford module bundle 
$\pi_{\mbox{\tiny$\mathcal{E}$}}:\,(\mathcal{E},\gamma_{\mbox{\tiny$\mathcal{E}$}})\rightarrow(M_1,g_1)$. Likewise, the bundle \eqref{clifftwistharmap} may be regarded as a twisted Clifford module bundle with the twisting given by the twisted Clifford module bundle
$\varphi^\ast\mathcal{E}_2\ot_{\mbox{\tiny$M_1$}}Cl_1\rightarrow(M_1,g_1)$.

A Clifford connection on \eqref{clifftwistharmap} reads 
\beq
\label{cliffcontwistharmap}
\partial_{\!\mbox{\tiny${\rm A}'$}} = \da\ot{\rm Id}_{\mbox{\tiny${\rm Cl}_1$}} +
{\rm Id}_{\mbox{\tiny$\mathcal{E}$}}\ot\nabla^{\mbox{\tiny${\rm Cl}_1$}}\,.
\eeq
Here,
\beq
\label{cliffcongen}
\da = \partial_{\!\mbox{\tiny${\rm A}_1$}}\!\ot{\rm Id}_{\mbox{\tiny$\mathcal{E}_2$}} +
{\rm Id}_{\mbox{\tiny$\mathcal{E}_1$}}\!\ot\nabla^{\mbox{\tiny$\varphi^\ast\mathcal{E}_2$}}
\eeq
denotes a general Clifford connection on \eqref{cliffharmap}, where, respectively,
$\nabla^{\mbox{\tiny${\rm Cl}_1$}}$ and $\nabla^{\mbox{\tiny$\mathcal{E}_2$}}$ are general
(Clifford) connections on the Clifford module bundles $Cl_1\rightarrow(M_1,g_1)$ and 
$\pi_2:\,(\mathcal{E}_2,g_2)\rightarrow(M_2,g_2)$. Clearly, on $Cl_1\rightarrow(M_1,g_1)$ there 
is a canonical choice provided by the Levi-Civita connection with respect to $g_1$.

Note that for all sections $\psi\in\goth{S}ec(M_1,\mathcal{E})$:
\beq
\partial_{\!\mbox{\tiny${\rm A}'$}}(\psi\ot 1) = \da\psi\ot 1\,.
\eeq

Let $e_1,e_2,\ldots,e_{n_1}\in\goth{S}ec(U_1,TM_1)$ be a local (oriented orthonormal) frame
on the open subset $U_1\subset M_1$ with the dual frame denoted as 
$e^1,e^2,\ldots,e^{n_1}\in\goth{S}ec(U_1,T^\ast\!M_1)$.

The quantization of \eqref{cliffcontwistharmap} then reads
\beq
\ddd_{\!\!\mbox{\tiny${\rm A}'$}} = \,\dda\ot{\rm Id}_{\mbox{\tiny${\rm Cl}_1$}} +
\sum_{a=1}^{n_1}\gamma_{\mbox{\tiny$\mathcal{E}$}}(e^a)\ot\nabla_{\!\!e_a}^{\mbox{\tiny${\rm Cl}_1$}}\,.
\eeq
It follows that $\ddd_{\!\!\mbox{\tiny${\rm A}'$}}(\psi\ot 1) = \dda\psi\ot 1$.

The Jacobi map of $\varphi$ can be identified with the section 
$d\varphi\in\Omega^1(M_1,\varphi^\ast TM_2)$.
We set for all $t\in U_1$: $\varphi_a(t) \equiv d\varphi(t)e_a(t)\in T_{\varphi(t)}M_2$, such that 
$d\varphi = \sum_{a=1}^{n_1} e^a\!\ot\varphi_a$ and consider
\beq
\label{chiharmap}
\begin{split}
\chi &= \sum_{a=1}^{n_1}e^a\ot\chi_a\in
\Omega^1(M_1,{\rm End}^-_{\mbox{\tiny$\gamma$}}(\mathcal{E}))\,,\\
\chi_a 
&:= \big({\rm Id}_{\mbox{\tiny$\mathcal{E}_1$}}\!\ot\gamma_2\big) 
\big({\rm Id}_{\mbox{\tiny$\mathcal{E}_1$}}\!\ot\varphi_a^\flat\big)\\
&= {\rm Id}_{\mbox{\tiny$\mathcal{E}_1$}}\!\ot\gamma_2(\varphi_a^\flat)\in
\mathcal{C}^\infty\big(U_1,{\rm End}^-_{\mbox{\tiny$\gamma$}}(\mathcal{E})\big)\,.
\end{split} 
\eeq
Accordingly, we set
\beq
\phi_{\mbox{\tiny D}} \;\stackrel{loc.}{:=}\;
\sum_{a = 1}^{n_1}{\rm Id}_{\mbox{\tiny$\mathcal{E}_1$}}\ot\gamma_2(\varphi_a^\flat)\ot e^a\in
\Omega^0(M_1,{\rm End}^-_{\mbox{\tiny$\gamma'$}}(\mathcal{E}'))
\eeq
and consider the following Dirac operator of simple type, which acts on sections of \eqref{clifftwistharmap}:
\beq
\label{dirharmapdop}
\DDD := \,\ddd_{\!\!\mbox{\tiny${\rm A}'$}} + \tau_{\!\mbox{\tiny$\mathcal{E}'$}}\,\phi_{\mbox{\tiny D}} \,.
\eeq

For sections $\psi \equiv \psi\ot 1\in\goth{S}ec(M_1,\mathcal{E}')$ one gets
\beq
\big<\psi,\,\DDD\psi\big>_{\!\mbox{\tiny$\mathcal{E}'$}} = 
\big<\psi,\,\dda\psi\big>_{\!\mbox{\tiny$\mathcal{E}$}}\,. 
\eeq
Furthermore,
\beq
\begin{split}
\|\,\phi_{\mbox{\tiny D}}\|^2 &\equiv 
\epsilon_1\sum_{a,b=1}^{n_1}g^\ast_1(e^a,e^b)\,tr_{\!\mbox{\tiny$\mathcal{E}$}}\chi^\dagger_a\chi_b\\
&= 
\pm\epsilon_1\epsilon_2 N\|d\varphi\|^2\,,
\end{split}
\eeq
where $\|d\varphi\|^2 \equiv \sum_{a,b=1}^{n_1}g_1^\ast(e^a,e^b)\,g_2|_\varphi(\varphi_a,\varphi_b) =
\sum_{a,b=1}^{n_1}g_1^\ast(e^a,e^b)\,\varphi^\ast\!g_2(e_a,e_b)$ and $N \equiv N_1+N_2$ is the rank
of the bundle \eqref{cliffharmap}\,.

We thus proved the following 

\begin{proposition}\label{dirharmact}
The total Dirac action \eqref{totdiract} with respect to the simple type Dirac operators \eqref{dirharmapdop}
reads
\beq
\label{dharmaction}
\begin{split}
\mathcal{I}_{\mbox{\tiny D,tot}}(\,\DDD,\psi) &=
\int_{M_1}\Big(\big<\psi,\,\DDD\psi\big>_{\!\mbox{\tiny$\mathcal{E}'$}} +
tr_{\!\gamma'}curv(\,\ddd_{\!\!\mbox{\tiny${A}'$}}) \pm \|\,\phi_{\!\mbox{\tiny D}}\|^2\Big)dvol(g_1)\\[.2cm]
&= 
\int_{M_1}\!\Big(-\epsilon_1\mbox{\small$\frac{rk(\mathcal{E}')}{4}$}scal(g_1) +
\big<\psi,\,\dda\psi\big>_{\!\mbox{\tiny$\mathcal{E}$}}  
\pm\epsilon_1\epsilon_2 N\|d\varphi\|^2\Big)dvol(g_1)\,.
\end{split}
\eeq
\end{proposition}

Therefore, up to the Einstein-Hilbert action (e.g. for fixed metric on $M_1$), the Dirac action 
basically coincides with the functional of non-linear $\sigma-$models (Dirac harmonic maps).
This holds true, especially, in the case $dim(M_1) = 2$.

\subsection{Geodesics as an example}
To present the archetype of a non-linear $\sigma-$model, let $(M_2,g_2) \equiv (M,g_{\mbox{\tiny M}})$ 
be an arbitrary $n-$dimensional smooth Riemannian manifold and $(M_1,g_1) := [0,1]\subset\rr^{1,0}$.  
For $(\mathcal{E}_1,\gamma_1)$ we set 
$\mathcal{E}_1 := [0,1]\times\hspace{-.25cm}\phantom{x}^{\mbox{\tiny 2}}\rr$ and consider 
the canonical Clifford map
\beq
\label{cliffactspin}
\begin{split}
\gamma_1:\,\rr &\longrightarrow \hspace{-.25cm}\phantom{x}^{\mbox{\tiny 2}}\rr\\
dt &\mapsto e\equiv(1,-1).
\end{split}
\eeq
Here, $\hspace{-.25cm}\phantom{x}^{\mbox{\tiny 2}}\rr\simeq Cl_{1,0}$ denotes the two-dimensional 
real algebra of {\it Study numbers} and $Cl_{1,0}$ is the Clifford algebra of the one-dimensional 
Euclidean space $\rr^{1,0}$. All trivial bundles are identified with their typical fibers, such that,
for instance, $T^\ast\!M_1$ is identified with $\rr$ etc..

We call in mind that the real algebra of Study numbers equals the two-dimensional real vector space $\rr^2$ with component wise multiplication. The unit is given by $1 \equiv (1,1)$, such that 
$\rr\hookrightarrow\hspace{-.25cm}\phantom{x}^{\mbox{\tiny 2}}\rr$ is contained as a canonical sub-algebra. 
Furthermore, $e\equiv(1,-1)\in\hspace{-.25cm}\phantom{x}^{\mbox{\tiny 2}}\rr$ is analogues to
$i\equiv(1,-1)\in\cc$. It follows that $\hspace{-.25cm}\phantom{x}^{\mbox{\tiny 2}}\rr\simeq_\rr{\rm End}(S\!\op\!\check{S})$, 
where $S := \{(u,0)\,|\,u\in\rr\}\subset\hspace{-.25cm}\phantom{x}^{\mbox{\tiny 2}}\rr\simeq Cl_{1,0}$ and 
$\check{S} := \{(0,v)\,|\,v\in\rr\}\subset\hspace{-.25cm}\phantom{x}^{\mbox{\tiny 2}}\rr\simeq Cl_{1,0}$
are the spinor modules with respect to the primitive idempotents 
$(1\pm e)/2\in\hspace{-.25cm}\phantom{x}^{\mbox{\tiny 2}}\rr$.

We put $(\mathcal{E}_2,\gamma_2) := (\Lambda T^\ast\!M,\gamma_{\mbox{\tiny Ch}})$, such that
\beq
\mathcal{E} = [0,1]\times\big(\,\hspace{-.25cm}\phantom{x}^{\mbox{\tiny 2}}\rr\ot\varphi^\ast\Lambda T^{\ast}\!M\big)
\simeq \varphi^\ast\Lambda T^{\ast}\!M\op\varphi^\ast\Lambda T^{\ast}\!M\,.
\eeq

Any section $\psi\in\goth{S}ec([0,1],\mathcal{E})$ thus corresponds to a pair of sections of the (trivial) algebra bundle
$\varphi^\ast\Lambda T^{\ast}\!M\simeq[0,1]\times\Lambda\rr^n\rightarrow [0,1]$.

Let again $I_k = (i_1,\ldots,i_k)$ be a multi-index for all $1\leq k\leq n = dim(M)$ and $i_l = 1,\dots,n\;(l=1,\dots,k)$. We make again usage of the shorthand
$\sum_I\sigma_I\equiv\sum_{k =1}^n\sum_{I_k}\sigma_{I_k}$. Then, $\psi(t) = \sum_I \psi_I(t)\ot e_I$,
whereby $e_{I_k}:\,[0,1]\rightarrow\varphi^\ast\Lambda T^\ast M,\;t\mapsto(t,{\bf e}_{I_k})$ are the canonical sections 
with ${\bf e}_{I_1},\ldots,{\bf e}_{I_n}\in\Lambda\rr^n$ being the standard basis. Furthermore, 
$\psi_{I_k}(t) = 
(\alpha_{I_k}(t),\beta_{I_k}(t))\in\hspace{-.25cm}\phantom{x}^{\mbox{\tiny 2}}\rr$ for all $k = 1,\ldots,n$.

We choose the trivial connection to define $\nabla^{\mbox{\tiny$\mathcal{E}_1$}}$, such that for all 
smooth sections
$\chi = (\chi_1,\chi_2)\in\goth{S}ec([0,1],\mathcal{E}_1)\simeq
\mathcal{C}^\infty([0,1],\hspace{-.25cm}\phantom{x}^{\mbox{\tiny 2}}\rr)$ the action of the corresponding Dirac operator reads: $\ddd_{\!\mbox{\tiny 1}}\chi(t) = (\dot{\chi}_1(t),-\dot{\chi}_2(t))\in\hspace{-.25cm}\phantom{x}^{\mbox{\tiny 2}}\rr$. 
Here, $\dot{\chi}_k(t) := d\chi_k(t)\partial_t|_t$, whereby  
$\partial_t:\,[0,1]\rightarrow[0,1]\times\rr,\;t\mapsto(t,1)$ is the canonical tangent vector field on 
$M_1 = [0,1]$. On the Clifford module  bundle $\Lambda T^\ast\!M\rightarrow M$ we take the induced 
Levi-Civita connection of $(M,g_{\mbox{\tiny M}})$.

With respect to our notation the action of the Dirac operator $\ddd$ explicitly reads:
\beq
\begin{split}
\ddd\psi(t) = \hspace{6cm}\\ 
\sum_I\!\Big(\dot{\alpha}_I(t) + \sum_{I'}\Gamma_{\!II'}|_{\varphi(t)}(\dot{\varphi}(t))\alpha_{I'}(t),\, -\dot{\beta}_I(t) - 
\sum_{I'}\Gamma_{\!II'}|_{\varphi(t)}(\dot{\varphi}(t))\beta_{I'}(t)\Big)\!\ot{\bf e}_I\,,
\end{split}
\eeq
 where for all $k = 1,\ldots,n$:
\beq
\sum_{l=1}^n\sum_{I_l}\Gamma_{\!I_kI_l}|_{\varphi(t)}(\dot{\varphi}(t))e_{I_l}(t) 
:= \nabla^{\varphi^\ast\!\Lambda T^\ast M}_{\!\!\partial_t}e_{I_k}(t)\,,
\eeq
defines the induced Levi-Civita connection coefficients of the pull-back connection and
$\dot{\varphi}(t) := d\varphi(t)\partial_t|_t\in T_{\varphi(t)}M$ is the velocity vector of the smooth curve 
$\varphi:\,[0,1]\rightarrow M$.

After appropriate normalization, the total Dirac action \eqref{totdiract} simplifies to
\beq
\label{susy1}
\begin{split}
\mathcal{I}_{\mbox{\tiny D,tot}}\big(\,\DDD,\psi\big) = 
\int_0^1\!\!\Big(\big<\psi,\ddd\psi\big>_{\!\mbox{\tiny$\mathcal{E}$}} + \|d\varphi\|^2\Big)dt\,.
\end{split}
\eeq
In particular, the universal Dirac action \eqref{stypediract} reduces to what is referred to as the 
{\it energy functional} of the curve $\varphi$:
\beq
\label{geod}
\begin{split} 
\mathcal{I}_{\mbox{\tiny D}}\big(\,\ddd + \tau_{\!\mbox{\tiny$\mathcal{E}'$}}\phi_{\mbox{\tiny D}}\big) =
\int_0^1g_{\mbox{\tiny M}}|_{\varphi(t)}\big({\dot\varphi}(t),{\dot\varphi}(t)\big)\,dt\,,
\end{split}
\eeq
where $d\varphi = dt\ot{\dot\varphi}\in\Omega^1\big([0,1],\varphi^\ast TM\big)$.

For fixed metric, the minima of \eqref{geod} are known to be given by the geodesics on 
$(M,g_{\mbox{\tiny M}})$.

\section{The Yang-Mills action}
A Dirac operator of simple type on an arbitrary Clifford module bundle \eqref{cliffmodbdl} is 
uniquely defined in terms of a Clifford connection $\da$ together with a section 
$\phi_{\mbox{\tiny D}}\in\goth{S}ec(M,{\rm End}^-_\gamma(\mathcal{E}))$. It is thus natural 
to consider Dirac operators of simple type, which are fully determined by Clifford connections. 
Of course, quantized Clifford connections $\,\DDD := \,\dda$ are special cases thereof. In this
case, the universal Dirac action \eqref{diract} reduces to the Einstein-Hilbert action. In this section 
we specify to Dirac operators of simple type which are fully determined by Clifford connections but
$\,\DDD \not= \,\dda$. The restriction of the universal Dirac action to these Dirac operators 
becomes the Yang-Mills functional.
 
Let $M_1 = M_2 \equiv M$ be a smooth orientable manifold of dimension $n\geq 1$. We assume $M$ 
to be closed compact. Also, let $g_1 = g_2 \equiv g_{\mbox{\tiny M}}$ be a (semi-)Riemannian metric of arbitrary signature. For $\varphi = {\rm Id}_{\mbox{\tiny M}}$ we denote by 
\beq
\label{twistingcurv}
F_{\!\mbox{\tiny A}} = F_{\!\mbox{\tiny${\rm A}_1$}}\ot{\rm Id}_{\mbox{\tiny$\mathcal{E}_2$}} +
{\rm Id}_{\mbox{\tiny$\mathcal{E}_1$}}\ot F_{\!\mbox{\tiny${\rm A}_2$}}\in
\Omega^2(M,{\rm End}^+_\gamma(\mathcal{E}))
\eeq
the {\it twisting (relative) curvature} on \eqref{cliffharmap} of the hermitian Clifford connection 
\beq
\da := \partial_{\!\mbox{\tiny${\rm A}_1$}}\!\ot{\rm Id}_{\mbox{\tiny$\mathcal{E}_2$}} +
{\rm Id}_{\mbox{\tiny$\mathcal{E}_1$}}\!\ot\partial_{\!\mbox{\tiny${\rm A}_2$}}\,.
\eeq

Let again $e_1,\dots,e_n\in\goth{S}ec(U,TM)$ be a local (oriented orthonormal) frame with dual frame
$e^1,\dots,e^n\in\goth{S}ec(U,T^\ast\!M)$. We set
\beq
\label{ymchi}
\begin{split}
\chi &\stackrel{loc.}{=} \sum_{a = 1}^n e^a\ot\chi_a\in\Omega^1(M,{\rm End}^-_\gamma(\mathcal{E}))\,,\\
\chi_a &:= \big({\rm Id}_{\mbox{\tiny$\mathcal{E}_1$}}\!\ot\gamma_2\big)
\big(int(e_a)F_{\!\mbox{\tiny A}}\big)\\
&=
\sum_{b=1}^{n_1}F_{\!\mbox{\tiny${\rm A}_1$}}(e_a,e_b)\ot\gamma_2(e^b) +
{\rm Id}_{\mbox{\tiny$\mathcal{E}_1$}}\!\ot F_{\!\mbox{\tiny${\rm A}_2$}}(e_a,e_b)\gamma_2(e^b)\in
\mathcal{C}^\infty(U,{\rm End}^-_\gamma(\mathcal{E}))\,.
\end{split}
\eeq
The reader may compare this with the case \eqref{chiharmap} of non-linear $\sigma-$models, 
whereby the correspondence is given by
$\varphi^\flat_a(v) \leftrightarrow F_{\!\mbox{\tiny A}}(e_a,v)$, for all $v\in TU$. In contrast to the
case \eqref{chiharmap}, however, the choice \eqref{ymchi} is most natural, as already mentioned, for 
it allows in a canonical way to also define the zero-order part $\phi_{\mbox{\tiny D}}$ of a simple type 
Dirac operator $\,\DDD$ in terms of the Clifford connection of $\,\DDD$. In this case one has
\beq
\|\,\phi_{\mbox{\tiny D}}\|^2 = \epsilon_1\epsilon_2\|F_{\!\mbox{\tiny A}}\|^2\,, 
\eeq
where 
\beq
\begin{split}
\|F_{\!\mbox{\tiny A}}\|^2 &\equiv 
-\sum_{a,b=1}^{n_1}g^\ast_{\mbox{\tiny M}}(e^a,e^c)\,g^\ast_{\mbox{\tiny M}}(e^b,e^d)\,
tr_{\!\mbox{\tiny$\mathcal{E}$}}\big(F_{\!\mbox{\tiny A}}(e_a,e_b)F_{\!\mbox{\tiny A}}(e_c,e_d)\big)\\ 
&\equiv
-\sum_{a,b=1}^{n_1}tr_{\!\mbox{\tiny$\mathcal{E}$}}F_{\!ab}F^{ab}\in\mathcal{C}^\infty(M)\,.
\end{split}
\eeq

We thus proved the following
\begin{proposition}
When restricted to the class of simple type Dirac operators considered, the universal Dirac action reads
\beq
\mathcal{I}_{\mbox{\tiny D}} =
\int_M\!\ast\Big(-\epsilon_1\mbox{\small$\frac{rk(\mathcal{E}')}{4}$}scal(g_{\mbox{\tiny M}}) 
%\big<\psi,\,\dda\psi\big>_{\!\mbox{\tiny$\mathcal{E}$}} 
+ \epsilon_1\epsilon_2\,\|F_{\!\mbox{\tiny A}}\|^2\Big)\,.
\eeq
\end{proposition}

\hspace{.1cm}

We finally discuss the case of {\it twisted spinor bundles}, usually encountered in the literature dealing 
with twisted spin Dirac operators. To this  end let $M$ be a {\it spin manifold} and 
$\pi_{\mbox{\tiny S}}:\,S\rightarrow M$ be a {\it spinor bundle}.  Moreover, let 
$\pi_{\mbox{\tiny E}}:\,E\rightarrow M$ be a smooth (odd) hermitian vector bundle. In this particular 
case, we set $\mathcal{E}_1 := S\ot_{\mbox{\tiny M}}E$ and $\mathcal{E}_2 := Cl_{\mbox{\tiny M}}$. 
The canonical embedding $\mathcal{E}_1\hookrightarrow\mathcal{E},\;z \mapsto z\ot 1$ then yields 
$\da(\psi\ot 1) = \partial_{\!\mbox{\tiny${\rm A}_1$}}\!\psi\ot 1$. The Clifford connection 
$\partial_{\!\mbox{\tiny${\rm A}_1$}}\equiv\nabla^{\mbox{\tiny$S\!\ot\!E$}}$ is but the 
{\it twisted spin connection} with the twisting curvature being given by 
$F_{\!\mbox{\tiny${\rm A}_1$}} = {\rm Id}_{\mbox{\tiny S}}\!\ot\!F^{\mbox{\tiny E}}$, 
where $F^{\mbox{\tiny E}}\in\Omega^2(M,{\rm End}^+(E))$ denotes the curvature of some (even)
hermitian connection $\nabla^{\mbox{\tiny E}}$ on $\pi_{\mbox{\tiny E}}:\,E\rightarrow M$.

When restricted to the sub-bundle 
$\pi_{\mbox{\tiny$\mathcal{E}$}}|_{\mathcal{E}_1}:\,\mathcal{E}_1\subset\mathcal{E}\rightarrow M$ 
the twisting curvature \eqref{twistingcurv} reduces to
\beq
\label{twistingcurvspin}
\begin{split}
F_{\!\mbox{\tiny A}} = {\rm Id}_{\mbox{\tiny S}}\!\ot\!F^{\mbox{\tiny E}}\!\ot{\rm Id}_{\mbox{\tiny Cl}}
=
F_{\!\mbox{\tiny${\rm A}_1$}}\!\ot{\rm Id}_{\mbox{\tiny Cl}}\,.
\end{split}
\eeq

Therefore, in the case of twisted spinor bundles the total Dirac action decomposes into the sum of the Dirac, the Einstein-Hilbert and the Yang-Mills action:
\beq
\label{DEHYM}
\begin{split}
\mathcal{I}_{\mbox{\tiny D,tot}} 
&=
\int_M\!\ast\Big(\big<\psi,\,\ddd_{\!\!\mbox{\tiny${\rm A}_1$}}\!\psi\big>_{\!\mbox{\tiny$\mathcal{E}_1$}} 
- \epsilon_1\mbox{\small$\frac{rk(\mathcal{E}')}{4}$}scal(g_{\mbox{\tiny M}}) 
+ \mbox{\small$2^n rk(S)$}\epsilon_1\epsilon_2\,\|F^{\mbox{\tiny E}}\|^2\Big)\\
&\equiv
\!\int_M\!\!\!\ast\big<\psi,\,\ddd_{\!\!\mbox{\tiny${\rm A}_1$}}\!\psi\big>_{\!\mbox{\tiny$\mathcal{E}_1$}}\!
 -
\epsilon_1\mbox{\small$\frac{rk(\mathcal{E}')}{4}$}\!\!\int_M\!\!\!\ast scal(g_{\mbox{\tiny M}}) +
\mbox{\small$2^{n+1}$}\epsilon_1\epsilon_2\!\!\int_M\!\!\!
tr_{\mbox{\!\tiny$\mathcal{E}_1$}}\!(F_{\!\mbox{\tiny${\rm A}_1$}}\!\wedge\ast F_{\!\mbox{\tiny${\rm A}_1$}})\,.
\end{split}
\eeq

\section{The Dirac-Harmonic-Yang-Mills action}
For further discussions, especially in the context of the Standard Model and gravity with torsion, we 
eventually present the setup that allows to combine the actions \eqref{dharmaction} and \eqref{DEHYM}.

For this let us consider the geometrical situation encountered in the case of non-linear $\sigma-$models.
We also assume that $(M_1,g_1)$ is a (closed compact) (semi-)Riemannian spin manifold of even dimension $n_1 \geq 2$ and
\beq
\label{twistbdl}
\pi_{\mbox{\tiny W}}:\,W\longrightarrow M_1
\eeq
a smooth (odd) hermitian vector bundle with grading involution $\tau_{\mbox{\tiny W}}\in{\rm End}(W)$
and hermitian product $\big<\cdot,\cdot\big>_{\!\mbox{\tiny W}}$.

With respect to a chosen spin structure we consider the twisted spinor bundle:
\beq
\label{twistspinbdl}
\pi_1\,: \mathcal{E}_1 := S\ot E_1\longrightarrow M_1
\eeq
where 
$\pi_{\mbox{\tiny$S$}}:\,S\rightarrow M_1$ is a corresponding (complex) spinor bundle and
\beq
\label{twistcliffbdl}
\pi_{\mbox{\tiny${\rm E}_1$}}:\,E_1 := Cl_1\ot W\longrightarrow M_1
\eeq
is the $W-$twist of the canonical Clifford module bundle 
$\pi_{\mbox{\tiny${\rm Cl}_1$}}:\,(Cl_1,\gamma_{\mbox{\tiny${\rm Cl}_1$}})\rightarrow(M_1,g_1)$. The Clifford action, grading involution and hermitian structure are denoted by
\beq
\begin{split}
\gamma_{\mbox{\tiny${\rm E}_1$}} &:= \gamma_{\mbox{\tiny${\rm Cl}_1$}}\ot{\rm Id}_{\mbox{\tiny W}}\,,\\
\tau_{\mbox{\tiny${\rm E}_1$}} &:= \tau_{\mbox{\tiny${\rm Cl}_1$}}\ot\tau_{\mbox{\tiny W}}\,,\\
\big<\cdot,\cdot\big>_{\!\mbox{\tiny${\rm E}_1$}} &=
\big<\cdot,\cdot\big>_{\mbox{\tiny${\rm Cl}_1$}}\big<\cdot,\cdot\big>_ {\mbox{\tiny W}}\,.
\end{split}
\eeq

Accordingly, the Clifford action, grading involution and hermitian product of \eqref{twistspinbdl} are 
given by
\beq
\begin{split}
\gamma_1 &:= \gamma_{\mbox{\tiny S}}\ot{\rm Id}_{\mbox{\tiny${\rm E}_1$}}\,,\\
\tau_1 &:= \tau_{\mbox{\tiny S}}\ot\tau_{\mbox{\tiny${\rm E}_1$}}\,,\\
\big<\cdot,\cdot\big>_{\!\mbox{\tiny$\mathcal{E}_1$}} &=
\big<\cdot,\cdot\big>_{\mbox{\tiny S}}\big<\cdot,\cdot\big>_ {\mbox{\tiny${\rm E}_1$}}\,.
\end{split}
\eeq

For a given connection $\nabla^{\mbox{\tiny${\rm W}$}}$ on \eqref{twistbdl} let 
$\nabla^{\mbox{\tiny${\rm E}_1$}} = \nabla^{\mbox{\tiny${\rm Cl}_1\!\ot\! W$}}$ be the induced Clifford connection on the $W-$twist \eqref{twistcliffbdl}. The twisting curvature reads
\beq
\begin{split}
F^{\mbox{\tiny${\rm E}_1$}} \equiv {\rm Id}_{\mbox{\tiny${\rm Cl}_1$}}\ot F^{\mbox{\tiny W}}\in
\Omega^2(M_1,{\rm End}^+_{\gamma}(E_1))\,,
\end{split}
\eeq
with $F^{\mbox{\tiny W}}\in\Omega^2(M_1,{\rm End}(W))$ being the curvature of 
$\nabla^{\mbox{\tiny${\rm W}$}}$.

On the Clifford extension $\pi_{\mbox{\tiny$\mathcal{E}'$}}:\,(\mathcal{E}',\gamma')\rightarrow(M_1,g_1)$ of the twisted Clifford module bundle
\beq
\label{twistedcliffbdl2}
\pi_{\mbox{\tiny$\mathcal{E}$}}:\, \mathcal{E} \equiv \mathcal{E}_1\ot\varphi^\ast\mathcal{E}_2\longrightarrow M_1
\eeq
 we consider the section
\beq
\phi_{\mbox{\tiny D}} := \phi_1 + \phi_2\in{\goth S}ec(M_1,{\rm End}^-_{\gamma'}(\mathcal{E}'))\,,
\eeq
where
\beq
\begin{split}
\phi_1 &:= \sum_{b=1}^{n_1} {\rm Id}_{\mbox{\tiny$\mathcal{E}_1$}}\ot\gamma_2(\varphi_b^\flat)
\ot \gamma_{\mbox{\tiny${\rm Cl}_1$}}(e^b)\in{\goth S}ec(M_1,{\rm End}^-_{\gamma'}(\mathcal{E}'))\\
\phi_2 &:= \sum_{a,b=1}^{n_1}{\rm Id}_{\mbox{\tiny S}}\ot\gamma_{\mbox{\tiny${\rm Cl}_1$}}(e^a)\ot 
F^{\mbox{\tiny W}}(e_a,e_b)\ot{\rm Id}_{\mbox{\tiny$\varphi^\ast\mathcal{E}_2$}}\ot \gamma_{\mbox{\tiny${\rm Cl}_1$}}(e^b)\\
&\hspace{.1cm}
+\sum_{a,b=1}^{n_1}{\rm Id}_{\mbox{\tiny S}}\ot\gamma_{\mbox{\tiny E}}(e^a)\ot 
\varphi^\ast\!F^{\mbox{\tiny$\mathcal{E}_2$}}(e_a,e_b)\ot \gamma_{\mbox{\tiny${\rm Cl}_1$}}(e^b)\\
&\equiv
\sum_{a,b=1}^{n_1}{\rm Id}_{\mbox{\tiny S}}\ot\gamma_{\mbox{\tiny${\rm Cl}_1$}}(e^a)\ot 
F^{\mbox{\tiny$W\!\ot\!\varphi^{\!\ast}\!\mathcal{E}_2$}}(e_a,e_b)\ot \gamma_{\mbox{\tiny${\rm Cl}_1$}}(e^b)
\in{\goth S}ec(M_1,{\rm End}^-_{\gamma'}(\mathcal{E}'))\,.
\end{split}
\eeq
Here, $F^{\mbox{\tiny$\mathcal{E}_2$}}\in\Omega^2(M_2,{\rm End}_\gamma^+(\mathcal{E}_2))$ is the curvature of some chosen Clifford connection $\nabla^{\mbox{\tiny$\mathcal{E}_2$}}$ of the Clifford module bundle $\pi_{\mbox{\tiny$\mathcal{E}_2$}}:\,(\mathcal{E}_2,\gamma_2)\rightarrow(M_2,g_2)$.

On the Clifford extension of the twisted Clifford module bundle \eqref{twistedcliffbdl2} we consider the 
{\it Dirac operator of simple type}
\beq
\label{tordirop3}
\DDD := {\,/\!\!\!\!{\nabla}}^{\mbox{\tiny$\mathcal{E}'$}} + 
\tau_{\mbox{\tiny$\mathcal{E}'$}}\phi_{\mbox{\tiny D}}\,.
\eeq
Here,
\beq
\nabla^{\mbox{\tiny$\mathcal{E}'$}} := 
\nabla^{\mbox{\tiny$\mathcal{E}\!\ot\!{\rm Cl}_1$}}
\eeq
is the induced {\it Clifford connection} on the Clifford extension of \eqref{twistedcliffbdl2} and
\beq
\begin{split}
{\nabla}^{\mbox{\tiny$\mathcal{E}$}} &:=
\nabla^{\mbox{\tiny$\mathcal{E}_1\!\ot\!\varphi^{\!\ast}\!\mathcal{E}_2$}}\,,\\
\nabla^{\mbox{\tiny$\mathcal{E}_1$}} &:= 
\nabla^{\mbox{\tiny${\rm S}\!\ot\!{\rm E}_1$}}\,. 
\end{split}
\eeq

When the embedding of Clifford module bundles (over the identity on $M_1$) is taken into account:
\beq
\begin{split}
S\!\ot\! E\equiv S\!\ot\!W\!\ot\!\varphi^\ast\mathcal{E}_2&\hookrightarrow\mathcal{E}\\
s\ot w\ot z &\mapsto s\ot 1_{\mbox{\tiny${\rm Cl}_1$}}\ot w\ot z\,,
\end{split}
\eeq
one gets the following

\begin{proposition}\label{DEHYM-action}
By an appropriate re-definition of  $\phi_1, \phi_2\in{\goth S}ec(M_1,{\rm End}^-_\gamma(\mathcal{E}'))$ and of $\psi\in{\goth S}ec(M_1,S\!\ot\!E)\subset{\goth S}ec(M_1,\mathcal{E}')$ the total Dirac action with respect to the Dirac operator \eqref{tordirop3} reads:
\beq
\label{totdiractionres1}
\begin{split}
\mathcal{I}_{\mbox{\tiny D,tot}}(\,\DDD,\psi) \;\thicksim\;&
\int_{M_1}\hspace{-.2cm}\ast\Big(\big<\psi,\,{/\!\!\!\!\nabla}^{\mbox{\tiny$S\!\ot\!E$}}\psi\big>_{\!\mbox{\tiny$S\!\ot\!E$}} + 
\mbox{\small$\frac{1}{2}$}g^{\mbox{\tiny$\ast$}}_1\!\ot\!g_2|_\varphi(d\varphi, d\varphi)\Big)\\[.1cm] 
&
-
\int_{M_1}\!\!\!
\Big(\epsilon_1\!\ast\! scal(g_1) - tr\big(F^{\mbox{\tiny E}}\wedge\ast F^{\mbox{\tiny E}}\big)\Big)\,.
\end{split}
\eeq
\end{proposition}

The first integral is but the Dirac harmonic action functional, where for all homogeneous 
elements $\alpha\ot u,\,\beta\ot v\in T^\ast_tM_1\ot T_{\varphi(t)}M_2$ and $t\in M_1$:
\beq
g^{\mbox{\tiny$\ast$}}_1\!\ot\!g_2|_\varphi(\alpha\ot u, \beta\ot v)|_t :=
g^{\mbox{\tiny$\ast$}}_1|_t(\alpha,\beta)g_2|_{\varphi(t)}(u,v)\,.
\eeq
That is,
\beq
\begin{split}
g^{\mbox{\tiny$\ast$}}_1\!\ot\!g_2|_\varphi(d\varphi, d\varphi)|_t &= \sum_{a,b=1}^{n_1}
g^{\mbox{\tiny$\ast$}}_1|_t(e^a,e^b)g_2|_{\varphi(t)}(\varphi_a(t),\varphi_b(t))\\ 
&=
\sum_{a,b=1}^{n_1}
g^{\mbox{\tiny$\ast$}}_1|_t(e^a,e^b)\,(\varphi^\ast\!g_2)|_t(e_a,e_b)\,.
\end{split}
\eeq

The second integral on the right-hand side of \eqref{totdiractionres1} is of Einstein-Hilbert-Yang-Mills type.
The functional \eqref{totdiractionres1} thus combines the previously discussed examples of 
non-linear $\sigma-$model and of Yang-Mills gauge theory. 

%Notice that
%\beq
%tr\big(F^{\mbox{\tiny E}}\wedge\ast F^{\mbox{\tiny E}}\big) \equiv
%tr\big(F^{\mbox{\tiny W}}\wedge\ast F^{\mbox{\tiny W}}\big) + 
%tr\big(F^{\mbox{\tiny$\varphi^{\!\ast}\!\mathcal{E}_2$}}\wedge\ast F^{\mbox{\tiny$\varphi^{\!\ast}\!\mathcal{E}_2$}}\big)
%\eeq

\Proof
Similar to the previous examples the proof of the Proposition \eqref{DEHYM-action} basically rests on
the general form \eqref{stypediract} the Dirac action takes with respect to Dirac operators of simple type and because of the decomposition
\beq
tr_{\!\mbox{\tiny$\mathcal{E}'$}}\phi_{\mbox{\tiny D}}^2 = tr_{\!\mbox{\tiny$\mathcal{E}'$}}\phi_1^2 +
tr_{\!\mbox{\tiny$\mathcal{E}'$}}\phi_2^2\,,
\eeq
which is straightforward to verify. 
\QED

We close with a remark on how the geometrical description of the {\it``Higgs field''} fits with 
the presented scheme. A more thorough discussion of this issue will be given in a forthcoming work 
when the Standard Model action is revisited within the geometrical setup of simple type Dirac operators 
as discussed here.

To geometrically interpret the smooth mapping $\varphi:\,M_1\rightarrow M_2$ of non-linear 
$\sigma-$models as a Higgs field, we assume that
\beq
\label{higgsbdl}
\begin{split}
\pi_{\mbox{\tiny${\rm M}_2$}}:\,M_2 &\longrightarrow M_1\\
 x &\mapsto t 
\end{split}
\eeq
is a smooth hermitian vector bundle with the fiber metric denoted by $\big<\cdot,\cdot\big>_{\!\mbox{\tiny${\rm M}_2$}}$. Accordingly, $\varphi\in{\goth Sec}(M_1,M_2)$ is supposed to 
be a section of \eqref{higgsbdl}. In the sequel we consider \eqref{higgsbdl} as a (real) vector bundle that is 
associated to a principal $G-$bundle $\pi_{\mbox{\tiny Q}}:\,Q\rightarrow M_1$. Here, the Lie group $G$ 
is supposed to be compact and semi-simple. Furthermore, we identify \eqref{twistbdl} with an associated (odd) hermitian vector bundle that carries a representation of $G$.

To define the Clifford module bundle $(\mathcal{E}_2,\gamma_2)\rightarrow(M_2,g_2)$ we remark that
a (hermitian) connection $\nabla^{\mbox{\tiny${\rm M}_2$}}$ on \eqref{higgsbdl} together with the metric $g_1$ on $M_1$ yields a metric $g_2$ on $M_2$. We then put 
$\mathcal{E}_2 := Cl_2\equiv Cl(TM_2,g_2)$, whereby $\gamma_2$ denotes the left regular representation of the Clifford algebra with respect to $g_2$ onto itself. 

Due to the construction of $g_2$, it follows that for all $v\in TM_1$
\beq
\begin{split}
g_2\big(d\varphi(v),d\varphi(v)\big) &= 
\big<\nabla_{\!\!v}^{\mbox{\tiny${\rm M}_2$}}\varphi,\nabla_{\!\!v}^{\mbox{\tiny${\rm M}_2$}}\varphi\big>_{\!\mbox{\tiny${\rm M}_2$}} + 
g_1(v,v)\\
&=
g_2\big(\nabla_{\!\!v}^{\mbox{\tiny${\rm M}_2$}}\varphi,\nabla_{\!\!v}^{\mbox{\tiny${\rm M}_2$}}\varphi\big) 
+ g_1(v,v)\,.
\end{split}
\eeq

Therefore,
\beq
\label{higgskinterm}
\|d\varphi\|^2 = \|\nabla^{\mbox{\tiny${\rm M}_2$}}\varphi\|^2 + {\rm dim}(M_1)\,.
\eeq
Notice that $\|\nabla^{\mbox{\tiny${\rm M}_2$}}\varphi\|^2$ is purely quadratic in the section 
$\varphi\in{\goth Sec}(M_1,M_2)$. Hence, by appropriate re-scaling of the sections $\phi_1$ and  
$\varphi$ on the right-hand side of \eqref{higgskinterm} one ends up with
\beq
\label{hkt}
\|d\varphi\|^2 = \|\nabla^{\mbox{\tiny${\rm M}_2$}}\varphi\|^2 + \Lambda\,,
\eeq
where $\Lambda > 0$ is a ``true'' cosmological {\it constant}.

When \eqref{hkt} is taken into account together with the canonical embedding
\beq
\begin{split}
S\!\ot\!W &\hookrightarrow \mathcal{E} = \mathcal{E}_1\!\ot\varphi^\ast\mathcal{E}_2\\ 
%= S\!\ot\!Cl_1\!\ot\!W\!\ot\varphi^\ast Cl_2\\
s\ot z &\mapsto s\ot 1\ot z\ot 1
\end{split}
\eeq
and the section $\phi_2$ is replaced by
\beq
\phi_2 := \sum_{a,b=1}^{n_1}{\rm Id}_{\mbox{\tiny S}}\ot\gamma_{\mbox{\tiny${\rm Cl}_1$}}(e^a)\ot 
F^{\mbox{\tiny W}}(e_a,e_b)\ot{\rm Id}_{\mbox{\tiny$\varphi^\ast\!Cl_2$}}\ot\gamma_{\mbox{\tiny${\rm Cl}_1$}}(e^b)\,,
\eeq
the total Dirac functional \eqref{totdiractionres1} becomes (again, after proper re-scaling)
\beq
\label{totdiractionres2}
\begin{split}
\mathcal{I}_{\mbox{\tiny D,tot}}(\,\DDD,\psi) \;\thicksim\;&
\int_{M_1}\hspace{-.2cm}\ast\Big(\big<\psi,\,{/\!\!\!\!\nabla}^{\mbox{\tiny${\rm S}\!\ot\!{\rm W}$}}\psi\big>_{\!\mbox{\tiny${\rm S}\!\ot\!{\rm W}$}} + \|\nabla^{\mbox{\tiny${\rm M}_2$}}\varphi\|^2 
%\mbox{\small$\frac{1}{2}$}\nabla^{\mbox{\tiny${\rm M}_2$}}\varphi\wedge\ast\nabla^{\mbox{\tiny${\rm M}_2$}}\varphi 
+ \|F^{\mbox{\tiny W}}\|^2
%tr\big(F^{\mbox{\tiny W}}\!\wedge\ast F^{\mbox{\tiny W}}\big)
\Big) \\[.1cm] 
&
+ \int_{M_1}\!\!\!\ast\Big(-\epsilon_1scal(g_1) + \Lambda\Big)\,.
\end{split}
\eeq
The total Dirac action thus describes a purely gauge coupled fermion and Higgs field similar to the Standard Model of particle physics including the natural appearance of the cosmological constant. In fact, the setup presented allows a geometrical interpretation of the cosmological constant 
in terms of the kinetic term of the Higgs (c.f. \eqref{hkt}), which thereby turns the Einstein-Hilbert 
action \eqref{eh} into \eqref{ehc}.

\section{Conclusion}
The Yang-Mills action and the functional of non-linear $\sigma-$models can be both described by 
Dirac operators of simple type. In this way one may say that they have the same ``square root'' and 
the underlying generic form of these actions is provided by the functional \eqref{stypediract} 
generalizing the Einstein-Hilbert action with a cosmological constant \eqref{ehc}. Indeed, the EHC
is shown to be induced from Dirac operators of simple type of the form \eqref{dirharmapdop}, where
the mapping $\varphi$ is geometrically interpreted as a section of an hermitian vector bundle
\eqref{higgsbdl}.

The decomposition of the universal Dirac action in terms of the fields defining a Dirac operator is 
similar to the decomposition of manifest supersymmetric actions in terms of the fields defining the underlying super-field. From this point of view one may argue that certain classes of Dirac operators 
will give rise to supersymmetric actions. Indeed, the functional of Proposition \eqref{dirharmact} is 
known to have a supersymmetric interpretation (for a survey of supergravity, we refer to 
\cite{Freedman et al}). Especially for $dim(M_1) = 2$, the supersymmetric interpretation of the functional 
of Dirac harmonic maps plays a basic role in the discussion of (super) Riemann surfaces (for an appreciable mathematical survey of this issue we refer to Sec. 2.4 in \cite{Jost}). This will be discussed 
in some detail within the context of geometrical torsion in a forthcoming work.

\bibliographystyle{amsplain}
\def\dbar{\leavevmode\hbox to 0pt{\hskip.2ex \accent"16\hss}d}
\providecommand{\bysame}{\leavevmode\hbox to3em{\hrulefill}\thinspace}
\providecommand{\MR}{\relax\ifhmode\unskip\space\fi MR }

\providecommand{\MRhref}[2]{%
  \href{http://www.ams.org/mathscinet-getitem?mr=#1}{#2}
}
\providecommand{\href}[2]{#2}

\end{document}